\documentclass{iopart}
\usepackage{graphicx}
\newcommand{\req}[1]{(\ref{#1})}
\newcommand{\be}{\begin{equation}}
\newcommand{\ee}{\end{equation}}
\newcommand{\bea}{\begin{eqnarray}}
\newcommand{\eea}{\end{eqnarray}}

\newcommand{\cro}[1]{\left[#1\right]}

\newcommand{\avg}[1]{\langle{#1}\rangle}
\newcommand{\ovl}[1]{\overline{#1}}
\newcommand{\BE}{\begin{eqnarray}}
\newcommand{\EE}{\end{eqnarray}}
\newcommand{\BEn}{\begin{eqnarray*}}
\newcommand{\EEn}{\end{eqnarray*}}
\newcommand{\barr}{\begin{array}}
\newcommand{\earr}{\end{array}}

\newcommand{\bit}{\begin{itemize}}      
\newcommand{\eit}{\end{itemize}}
\newcommand{\bc}{\begin{center}}
\newcommand{\ec}{\end{center}}
\newcommand{\ben}{\begin{enumerate}}    
\newcommand{\een}{\end{enumerate}}

\newcommand{\erf}{{\rm erf}\,}

\begin{document}

%@twocolumnfalse\endcsname
\title{Exclusion particle models of limit order financial markets}
\author{Damien Challet and Robin Stinchcombe}
\address{Theoretical Physics, 1 Keble Road, Oxford OX1 3NP, United Kingdom}
\date{\today}
\ead{challet@thphys.ox.ac.uk, stinch@thphys.ox.ac.uk}
%\maketitle
%\widetext
%%%%%%%%%%%%%%%%%%%%%%%%%%%%%%%%%%%%%%%%%%%%%%%%%%%%%%%%%%%%%%%%%%%%%%%%%%%%%

\begin{abstract}Using simple particle models of limit order markets,
we argue that the mid-term over-diffusive price behaviour is inherent
to the very nature of these markets. Several rules for rate changes
are considered. We obtain analytical results for bid-ask spread properties, Hurst plots
and price increment correlation functions.\\
\end{abstract}
%]
%\pacs{}
%\narrowtext
\section{Introduction}

While physicists involved in academic research on financial markets
have focused on the analysis and modelling of 
time series such as stock market prices or
indices~\cite{Bouchaud,MantegnaStanley,Daco}, limit order markets are
addressed by economists in terms of market micro-structure~\cite{Review}.
 In such markets, traders can publish their wish to buy/sell a given
quantity of stock at a given price, that is, place a limit order. Most of the time, there is no
order of the opposite type at that price, and the limit order stays in
the order book until it is cancelled or matched by another order 
of the opposite type at the same price. An order that causes an
immediate transaction, as in the last example, is called market order.
One of the most interesting aspects of these markets is the interplay
between price dynamics and  order
dynamics. Whereas economists have studied and characterised many important aspects of these
markets~\cite{Review}, such as why traders would like to use
them after all~\cite{why}, and produced models with stochastic order
placement (see for instance~\cite{Domowitz,Cohen}),
 physicists prefer to consider orders as particles,
thinking of size as mass and price as spatial position; the price
evolution merely results from stochastic dynamics: it follows a
kind of random walk, hence is not
assumed to be fixed or bounded in contrast with other
models~\cite{Mendelson,Cohen,Domowitz}.  Data analysis performed
in  \cite{Coppejans,Maslov2,CS01,BouchaudLimit,FarmerLimit2}
shows that in the language of physics, limit order markets are
out-of-equilibrium, that is, there are similar to a bucket of water
heated by a campfire and filled by an irregular shower. There are two families of nonequilibrium particle models of
such markets,  one that assumes that orders diffuse on the price axis
\cite{BPS,Kogan,Gunter} and another one where orders can be placed, executed \cite{Domowitz,Maslov} and cancelled
\cite{CS01,FarmerLimit,BouchaudLimit}. There is evidence from market
data that orders do not diffuse on the
price axis, but are placed, and then executed or cancelled
 with time dependent rates \cite{CS01}. This begs for a
non-equilibrium model made up of these three processes
\cite{CS01,FarmerLimit,BouchaudLimit}. So far, all models have
assumed constant rates. Here we show that this assumption is
responsible for very unrealistic price behaviour, and that lifting this
assumption leads to the correct behaviour.

It is well documented that stock and foreign exchange prices are
over-diffusive over several months~\cite{Plerou,Daco}:
if the (log)-price $p_t$ followed
a standard random walk, its increments (or returns) would be such that
$p_\tau-p_0\sim \tau^H$ with $H=1/2$; $H$ is called the Hurst
exponent. Generally speaking, $H$ depends on $\tau$. For instance, real prices are reportedly
over-diffusive ($H>1/2$) for several months, and then tend to be
diffusive ($H=1/2$) for larger time lags~\cite{Plerou,Daco}. Over-diffusion implies some
sort of long-term memory. This behaviour is not
captured in the current particle models. On the contrary, $H=1/4$
for all times in models where no cancellation of orders is allowed~\cite{BPS,Kogan,Maslov}. 
It has been realised that order cancellation is
needed in order to obtain a crossover to diffusive price for
large time lags \cite{CS01}. On the other hand, over-diffusive behaviour is
obtained when price trends are introduced, that is, when the price has
a temporary tendency to rise or decrease, as in the
 model of Ref.~\cite{Gunter}, which has $H=2/3$ for all times,
when no order cancellation is allowed, and a crossover to $H=1/2$ when it is.
 Here we generalise the notion of trends introduced in
\cite{Gunter}, and study their influence on the Hurst exponent and the
return correlation function.

Nonequilibrium particle models are notoriously hard to tackle
analytically, and limit order market models are no exception.
Therefore one ought to study first the simplest models of this kind,
namely exclusion models \cite{Gunter}, where at most
one order is present at each tick. Recent work in 
nonequilibrium statistical mechanics shows that some of these particle models can be
exactly solved~\cite{Derrida} (see~\cite{Stinchcombe} for a
review). Exclusion is naturally found when there are so few orders
that it is unlikely to find two orders at the same tick, which is also
the limiting case where Ref~\cite{FarmerSupriya} obtained
 analytical results concerning order densities in a class of models introduced
in~\cite{CS01,FarmerLimit}. Beyond this seductive mathematical aspect,
exclusion models are arguably relevant for the modelling of limit
order market. Indeed, priority of execution is given to orders first
by price and then by chronological order of their deposition. This is
particularly important near the best prices, when impatient traders
are likely to be willing to take precedence over their colleagues. 
Therefore exclusion models of limit order
markets capture at least some part of the dynamics near the bid-ask spread,
which plays a major role in these markets, particularly when there is
no market maker. Another interpretation of the exclusion condition is
that it encodes the presence or absence of quotes at a given tick and
neglects the actual number of shares.
%Therefore,
%by `rotating' all orders, one can transform a
%given order distribution to an exclusion one, following
%Ref.~\cite{Supriya}, where an order of size $S$ occupies $S$ sites. 
%The exclusion process is particularly important near the best offers: if
%one is willing to have a priority over previous
%orders, one has to place a new order at a
%price where no order already stands, since orders are executed by
%chronological order. 

In short, the exclusion
hypothesis is a
way to reduce the dimensionality, hence the complexity, of limit order
market models, while keeping their collective character. Here we
transform the particle model found in Ref.~\cite{CS01} by identifying
processes in real markets into an exclusion model.

\section{Definition of the model}

Two types of orders, ask ($A$) and bid ($B$), also denoted by $-$ and
 $+$ and corresponding respectively to
 sell and buy orders, live on a semi-infinite line --- the
price whose unit is the tick.  As limit orders are a way of storing demand and offer, it
 is clear that all bid orders are found on the left of the ask orders. 
At time $t$, there are $n_i(t)\in\{-1,0,1\}$ orders on site
$i$, where $-1$ denotes the presence of an ask order and $1$ the
 presence of a bid order.  We denote by $b(t)$ the position of best
 bid price and $a(t)$ the position of the best ask price. The last paid price $p(t)$ is the
 position where a transaction last took place.  All quantities of
 interest depend on these last three
quantities. For instance, the bid-ask spread is the gap that
separates the best bid from the best ask; here we denote it by 
$s(t)=a(t)-b(t)-1$. We do
not consider any mechanism which would ensure the positivity of
the price, since we aim to study normal price dynamics; we argue that
special mechanisms arise in real markets when the price reaches very low
prices.\footnote{There is technical reason to justify this
assertion at least on the NASDAQ: a stock is unlisted from the NASDAQ
 if its price stays for too
long under 1\$ (see www.nasdaq.com). This is a motivation for the company to weigh on its
stock price.} 
We assume that all processes occur relative to the current best prices, that
 is, relative to $a(t)$ for asks and $b(t)$ for bids.
At each time step $t$, three different events can happen: a new order
of each type (bid+ask) is placed, a market order of each type is
 placed, and some orders are cancelled. Mathematically,
\ben
\item with probability $\delta_a$, a new ask order is placed into the
bid-ask spread. Its price $a(t)-\Delta$ ($\Delta>0$)
is randomly drawn according to some distribution probability
$P_a(\Delta)$. With probability $\delta_a'$, another ask order is
deposited on the right of the current best price at $a(t)+\Delta'$,
where $\Delta'>0$ is randomly drawn from $P_a'(\Delta')$. Idem for
bids: an new order is placed in the spread with probability $\delta_b$
at $b(t)+\Delta$, where $\Delta$ is taken from $P_b(\Delta)$. With
probability $\delta_b'$, a new order is placed at a lower price that
$b(t)$, that is, at $b(t)-\Delta'$ where $\Delta'$ is drawn from the
probability distribution function (pdf) $P_b'(\Delta')$. There can be
at most one order per site, hence,
if there is already a particle at $b(t)-\Delta'$, or at
$a(t)+\Delta'$, nothing happens.

\item Each particle has the same fixed probability $\eta$ to be cancelled.

\item With probability $\alpha_a(t)$ a market order eats the best ask order
located at $a(t)$ and with probability $\alpha_b(t)$
the bid order at $b(t)$ is executed.

\item Rates change according to predefined rules (see below).
\een

Deposition and market orders in particle limit order market models can be
 found in Ref.~\cite{Maslov},
cancellation was introduced in Ref.~\cite{CS01} and in
 Ref.~\cite{Cohen} in a different context; its main role is to
ensure that the total number of orders does not diverge with time.
 Note that $P(\Delta)$
has a major influence on the bid-ask spread dynamics;  the average
typical deposition position conditional to the bid-ask spread depends linearly on the
spread in real markets~\cite{CS01}, which we denote by
$\avg{\Delta^2|s(t)}^{1/2}=c_0+c_1 s(t)$, where $\avg{.}$ denote a
 temporal average.
This can be explained for instance by buy orders that are placed
at $b(t)+s(t)/2$, or very close to the best ask order, e.g. at
$a(t)-1$. Playing with $c_0$ and $c_1$ modifies the spread
dynamics. For instance, if $c_0>0$, $c_1=0$, there is no feed-back of the
spread on the width of deposition; if $c_1>1$, the contribution of
cross-deposition to market orders cannot be neglected. Finally,
if $c_0=0$ and $c_1>0$, the cross-deposition rate is constant.

 Several simplifying hypotheses will be made here: we will assume that
all deposition pdfs are equal, that is, $P_a=P_b=P_a'=P_b'$, and
Gaussian. This contrasts with empirical facts that $P'(\Delta)$ is a power-law over
two decades of ticks in the Paris Bourse~\cite{BouchaudLimit} in the London Stock
~\cite{FarmerLimit2}, and in the NASDAQ~\cite{Potters}, which
implies that the probability of far deposition is non-negligible.
 In addition, the cancellation
rate of a given order has a power-law dependence 
on its position relative to the best order, $\eta(\Delta)\sim
\Delta^{-\nu}$ with $\nu\simeq 2$~\cite{Potters}, which is consistent with the
functional form of the lifetime distribution previously
reported~\cite{Lo,CS01}: here, this dependence is neglected.

In real markets, rates are not fixed, but vary wildly during any
trading day. Moreover, we see no practical reasons why the rates
should obey a balance equation such that the expected
price drift is zero at all times, as it is assumed in all current models, except
that of Ref.~\cite{Gunter}. On the contrary, heterogeneity of market participants
is probably enough to argue that rates should be balanced only on average. As a first approximation,
we assume here that they are independent from each other. As we shall see, this
provides a universal mechanism for mid-term over-diffusive prices.
We study several alternative simple rules that stipulate how to change the rates, depending on the
quantity to be computed:
\ben
\item At each time step, with probability $p$, all rates are redrawn independently
at random. The Hurst exponent is then very easy to compute.
\item At each time step, each rate is redrawn with probability
$p$. This is also appropriate for the computation of the Hurst exponent.
\item Rates are linearly cross-correlated, that is,\bea\label{ratespm}
\alpha_a(t)=\alpha_0+\alpha_1\sigma(t) &~~~& \alpha_b(t)=\alpha_0-\alpha_1\sigma(t)\\
\delta_a(t)=\delta_0-\delta_1\sigma(t) &~~~& \delta_b(t)=\delta_0+\delta_1\sigma(t),
\eea
where $\sigma(t)=\pm 1$ is the only quantity that is changed with
probability $p$ at each time step. It defines the direction of 
trend.  This setup makes it easy to
compute the autocorrelation function and the properties of the bid-ask spread.
\item Same as rule (iii), with the additional constraint that
$\alpha_0=\alpha_1$ and $\delta_0=\delta_1$. In this case, market
orders occur at one side, and order deposition at the other.
\item Same as rule (iv), but the direction of the trend $\sigma(t)$ is
allowed to change only when the bid-ask spread is zero~\cite{Gunter}.
\een

In rules (i-iv) the probability of rate
change at each time step is constant. This sets the typical lifetime of a set
of rates to $1/p$, and that rates are exponentially
cross- and auto-correlated, whereas this dependence should be
algebraical~\cite{CS01,Stanleyvol}. Rule (v) relates trend changing to
the actual properties of the bid and ask distribution. Other similar
rules could also be studied, for instance by relating
probabilistically the trend to volume imbalance between the two
distributions. However, rule (v) is most appropriate for mathematical analysis.

\section{Bid-ask spread properties}
\label{bidaskspread}
Here we derive basic results about the bid-ask spread. We assume
that the width of order deposition does not depend on the bid-ask
spread ($c_1=0$), and that the price of new orders new prices $a(t)-1$ and
$b(t)+1$, that is, $P(\Delta)=\delta_{\Delta,1}$ where $\delta$ is the
Kronecker index. Finally, we neglect
cancellation, whose role is less crucial in exclusion models, as
the exclusion condition itself limits the total number of orders. The
last two assumptions imply that after a transient~\cite{transient},
the order density is uniform and equal to
$1$ for $p<b(t)$ and $p>a(t)$.

As long as the rates are fixed, the best ask
evolution is the following: $P[a(t+1)=a(t)+1]=\alpha_a$, and
$P[a(t+1)=a(t)-1]=\delta_a$, hence $a(t)$ follows a biased random walk
(RW) with
drift $\alpha_a-\delta_a$. The same applies to $b(t)$, with a drift of
$\delta_b-\alpha_b$, hence the mid-price, defined as
$m(t)=(a(t)+b(t))/2$, diffuses with an average drift
$(\alpha_a+\delta_b-\alpha_b-\delta_a)/2$. The evolution of the
spread $s(t)$ is also a biased RW, with drift
$v=\alpha_a+\alpha_b-\delta_a-\delta_b<0$. Since $s(t)\ge0$, the spread
undergo a biased RW with reflective boundary at $s=0$. Several
properties of $s$ can then be easily computed. For instance the pdf of the time during which $s>0$
is given by the first passage at $s=0$ of a particle
starting at $s_0$~\cite{Redner}
\be\label{F0t}
F(t)=\frac{s_0}{\sqrt{4\pi Dt^3}}\exp\cro{-\frac{(s_0+vt)^2}{4Dt}}.
\ee
where $D$ is the diffusion constant. In particular, when $v\to0_{-}$, $F(0,t)\simeq
t^{-3/2}$. For any
$v<0$, it is well known that the pdf $P(s,t)$ of random walks with
reflective boundary tends to a stationary state that is exponentially decreasing
\be
P(s)=\frac{|v|}{D}\,\exp(-|v|s/D)
\ee
and $\ovl{s}=D/|v|$, where $\ovl{.}$ denotes an average over $P(s)$.\footnote{In continuous space and time, $P(s,t)$ is indeed the solution of the
diffusion equation $D\partial^2 P/\partial s^2= -|v|\partial
P/\partial t$.} As shown by these equations, the spread properties only depend
on $v$ and $D$, provided that $v$ and $D$ remain constant long enough.

Rules (iii), (iv) and (v) for rate change are such that the change of trend
$\sigma(t)$ has no influence on the dynamics of $s(t)$, as $v$ and $D$
are independent of $\sigma(t)$, therefore the above equations are
always valid in this case. In particular, rule (iv) makes
the spread dynamics  a one-step process for $s>0$:
\bea\label{gap1step}
\frac{dP(s)}{dt}&=&[(1-\delta)(1-\alpha)+\alpha\delta]P(s)\nonumber\\&&+\delta(1-\alpha)P(s+1)+(1-\delta)\alpha P(s-1)
\eea
Note that for $s=0$, any deposition is in fact a cross-deposition,
hence the effective annihilation rate $\alpha_{\rm
eff}$ equals $\alpha+\delta$. Eq. \req{gap1step} shows again that the spread process is equivalent to a
biased RW with reflective boundary at $s=0$; its
 drift is $v=\alpha-\delta<0$, and its diffusion constant  $D=[\alpha(1-\alpha)+\delta(1-\delta)]/2$.
If
order cancellation is allowed, the 0-th order correction to the above
picture is $\delta\to\delta-\eta$ and
$\alpha\to\alpha+\eta$, hence $v\to v-2\eta$.

\begin{figure}
\centerline{\includegraphics[width=8cm]{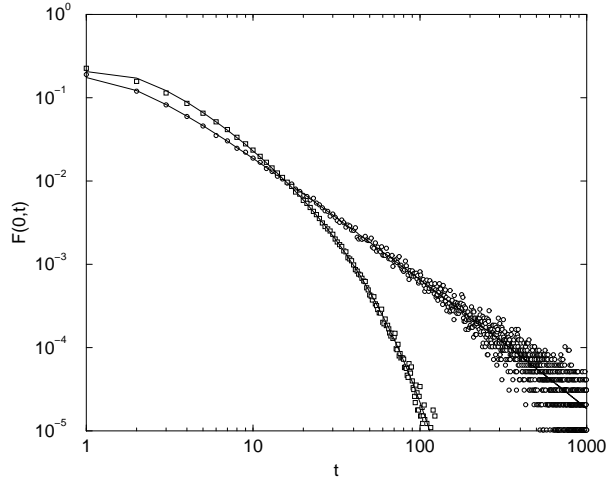}}
\caption{Trend duration pdf $F(0,t)$, for rule (iv)  ($s_0=1$, $\delta=0.5$, $\alpha=0.20$ (squares) and
$\alpha=0.48$ (circles)). $10^7$ iterations. Continuous lines are
theoretical predictions from Eq \req{F0t}.}
\label{flight}
\end{figure}

\section{Hurst exponent}

The Hurst exponent $H$ depends in general of $\tau$ in our
model. Irrespective of the considered rule for the update of the
rates, the price is over-diffusive ($H>1/2$) for intermediate $\tau$ and
diffusive ($H=1/2$) for large $\tau$. For rules (i)-(iv), the crossover
between $H>1/2$ and $H=1/2$ occurs for $\tau\sim 1/p$. When rule (v)
is applied, it occurs for $\tau$ of the order $\int dt F(0,t) t$.
The small $\tau$ behaviour depends on the rule and on the parameters,
as discussed below. 
% If the spread is
%typically small, $H=1/2$. If the spread is large on average, the price
%is diffusive for very small $\tau$, and then underdiffusive (see
%Fig.~\ref{over} and ~\ref{Hurst2an}

\subsection{Rule (i)}

\begin{figure}
\centerline{\includegraphics[width=8cm]{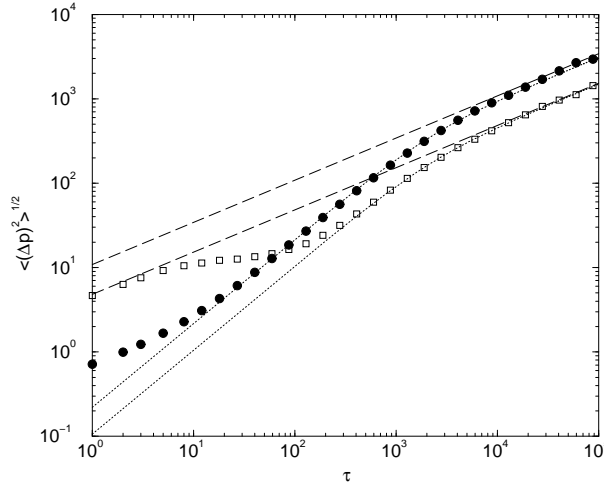}}
\caption{Hurst plot of the price increments for rule (i).  
. Market order rates drawn uniformly between 0 and 0.1, deposition
rate between 0 and 1 (circles), 0 and 0.5 (squares). $p=0.001$, $10^6$
iterations, no order cancellation. The dashed line
 represents normal random walk ($H=1/2$) and dotted lines are obtained from Eq.~\req{H}.} 
\label{over}
\end{figure}

The simplest setup for computing the Hurst exponent $H(\tau)$, is to
change the rates according to rule (i). For a given set of rates
$\alpha_a$, $\alpha_b$, $\delta_a$, $\delta_b$, the average speed of
the mid-price $m(t)$ is
$v=\alpha_a+\delta_b-\alpha_b-\delta_a$. Neglecting the diffusion of the
mid-price, and
retaining only its ballistic motion, one readily computes
$\avg{r(t+\tau)r(t)}\sim (1-p)^\tau$. Therefore, after a summation,
\be\label{H}
\avg{(p_{t+\tau}-p_t)^2}=\avg{(\sum_{t'=1}^\tau r_{t+t'})^2}\sim \tau+\frac{2(1-p)}{p^2}\,\cro{p\tau-1+(1-p)^\tau},
\ee
Figure~\ref{over} shows that Eq~\req{H} describes adequately the price
behaviour for large $\tau$. The discrepancy observed for small $\tau$
is probably due to bid-ask spread bouncing and other diffusive
behaviours that we neglected. Note that when the deposition rates are
close on average to the market order rates, a sub-diffusive behaviour appears for
small times. This general behaviour is due to spread bouncing,
as in this case, the spread is typically large. It appears for
all rules where frequent spread bouncing is allowed, owing to the
possibility of having simultaneous market orders at both sides, that is, rules (i),
(ii) and (iii).

\begin{figure}
\centerline{\includegraphics[width=8cm]{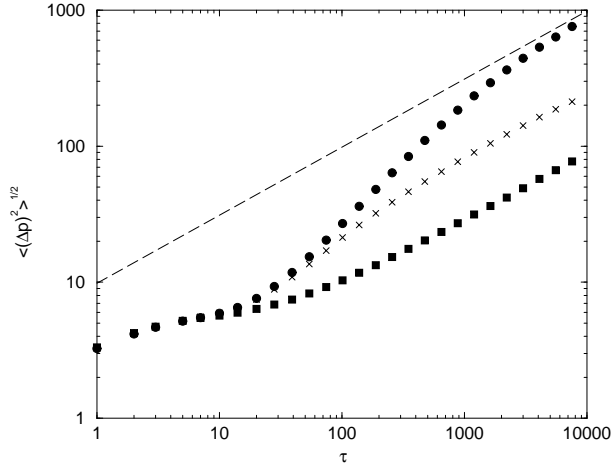}}
\caption{Hurst plot for rule (iii), showing
 characteristic under-diffusive prices at short
times. ($\delta=0.5$, $\alpha_0=0.22$, $\alpha_1=0.02$,
$\eta=0$, $p=0.001$, $0.01$ and $0.1$, $10^6$ iterations)}
\label{Hurst2an}
\end{figure}

\subsection{Rule (ii)}

Rule (ii) is arguably more realistic than rule (i). While it leads
to essentially the same behaviour, the calculus is more
complicated and of no special interest for our purpose.

\subsection{Rule (iii)}
\begin{figure}
\centerline{\includegraphics[width=8cm]{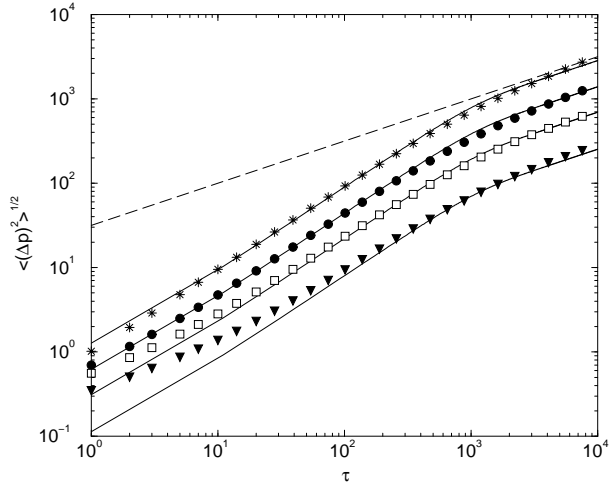}}
\caption{Hurst plot for rule (iv) for $\alpha=0.9$, $\delta=1$
 (stars), $\alpha=0.44$, $\delta=0.5$ (circles), $\alpha=0.22$,
 $\delta=0.25$ (squares), and $\alpha=0.08$, $\delta=0.1$ (triangles)
 ($p=0.001$, $\eta=0$ and $10^6$ iterations in all datasets). }
\label{Hurst1an}
\end{figure}

Fig.~\ref{Hurst2an}  shows that,
similarly to rule (i) and (ii),
if $p$ is small enough, the price is
under-diffusive, then over-diffusive, and tends to be diffusive at large
times. Since the
duration of the over-diffusive part is related to $1/p$, when $p$ is
too large, the price cannot develop its over-diffusive behaviour. Note
that the crossover from $H>1/2$ to $H=1/2$ has the same characteristic
form as in Figs~\ref{over} and~\ref{Hurst1an}. 
%This figure also confirms that spread bouncing is the cause of the excess diffusion also seen in Fig.~\ref{over}

\subsection{Rule (iv)}

In the case of Rule (iv), the price follows a persistent
RW~\cite{PRW}, that is, the probability that $r(t+1)=r(t)$ is
equal to $1-p$, where $p$ is the probability of trend reversal.
Therefore, the over-diffusive behaviour of our model can be qualitatively explained by mapping
the price increments ($\pm 1$) to an unidimensional (nearest
neighbour) Ising spin model where the space plays the role
of the time. A spin is a variable that can take two
values $-1$ and $+1$ (think of a price increment). The unidimensional
Ising model consists in placing a given number of these on a line, and to
assuming that spin at site $i$ is only influenced by its neighbours,
that is, the spins at sites $i+1$ and $i-1$ (see~\cite{Ising} for more
information). The mapping is possible, because the market model we
consider is Markovian. Generalising known results for such models to our case
where the price jumps over the spread when the trend reverses, whose effect
turns out to be negligible for large $\tau$, we find when $p$ is small and $\avg{s}\ll
1/p$ that for large $\tau$,
\be
\avg{[p(t+\tau)-p(t)]^2}\simeq\alpha^2\cro{\tau+\frac{\tau (1-p)}{p}\tanh[\tau
p/(1-p)]}~~~\tau\gg 1
\ee
which indicates a continuous crossover between $H(\tau)=1$ for small $\tau$ and
$H=1/2$ for large $\tau$. However, the crossover appears visually to occur in
a small time window.
The mapping between the market and Ising models gives good results,
even for  small $\tau$.

\subsection{Rule (v)}

This rule also gives rise to a persistent RW, hence to the same results as
those of rule (iv).

\section{Price autocorrelation}

In this section, we study how the price increments autocorrelation
function $C(\tau)=\avg{r(t+\tau)r(t)}$ depends on the rule of rate
changes and on the chosen parameters.

\subsection{Rule (iv)}
\label{subsect:p}
\begin{figure}
\centerline{\includegraphics[width=8cm]{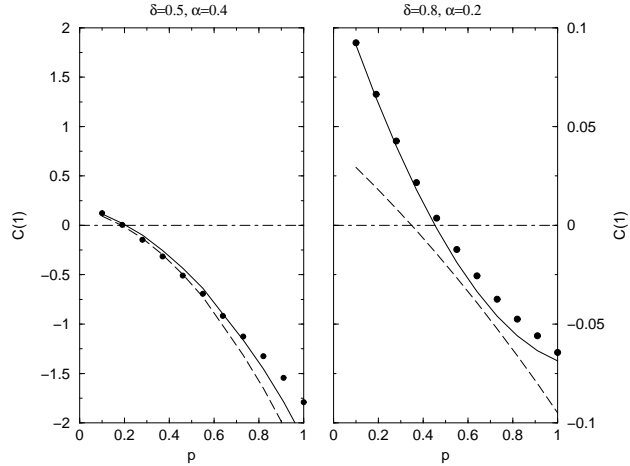}}
\caption{Price increment autocorrelation function $C(1)$ of
the model of subsection~\ref{subsect:p} (left panel: $\delta=0.5$,
$\alpha=0.4$; right panel: $\delta=0.8$, $\delta=0.4$. $\eta=0$, 
 average over $50$ runs, 10$^5$ iterations; continuous lines are
theoretical predictions of Eq~\req{C1}, and dashed lines are those of Eq.~\req{C1pjump}).}
\label{figC1pjump}
\end{figure}

Suppose that the probability of trend reversal is equal to $p$ at each time
step.  Neglecting the effect of cross-deposition when $s=0$, we find
\bea\label{C1pjump}
C(1)\simeq C_0(1)&=&\alpha^2[(1-p)^2-p(1-p)(1-\delta)\nonumber\\&& -p^2(\avg{s^2}+\avg{s}(1-\delta))].
\eea
As soon as the influence of spread bouncing is greater than that of annihilation,
$C(1)<0$. This is to be related to one known possible cause for negative
$C(1)$, spread bouncing \cite{Roll,Daco}, although here it only happens
whenever the trend $\sigma(t)$ changes.
When $P_s(0)$ is not negligible, Eq \req{C1pjump}
must be corrected: a transaction takes place if a market order is
placed ($\alpha$), or a market order is deposited onto the best
opposite order ($\delta$) (cross-deposition). If both happen during a time-step, we assume
that annihilation occurring before and after deposition are equally likely. We finally find
\bea\label{C1}
C(1)&\simeq&(1-P_s(0))C_0(1)\nonumber\\
&&+P_s(0)\{(1-p)^2\alpha[\alpha+\delta(1-\alpha/2)(1+\delta/2)]\nonumber\\
&&+p(1-p)[\alpha^2\delta/2-\alpha(\alpha+\delta+\alpha\delta^2/4)]  -p^2\alpha\delta^2\}
\eea
Figure~\ref{figC1pjump} shows that numerical simulations are in good
agreement with Eq~\req{C1}. Note that for this figure we used values of $P_s(0)$, $\avg{s}$
and $\avg{s^2}$ directly taken from the simulations, as the continuous
approximation for $P_s$ breaks down for typically small spreads. In the
right panel, the typical spreads is very small and $C_0(1)$ is not a good
approximation to $C(1)$. Eq~\req{C1} relates the price increments autocorrelation
function to the average spread and the second moment of the spread,
hence, can be seen as a generalisation of the seminal result of
Ref~\cite{Roll}, and is related other works~\cite{Madhavan,Hull}.

\subsection{Rule (iii)}

Simultaneous transactions on both sides of the market are an additional source for short term
negative return autocorrelation. Results of the previous subsection can be
generalised to this case: the spread dynamics is described by a two-step
process, which is again equivalent to a biased random walk with
reflective boundary at $s=0$, and $C(1)$ can be calculated, but leads
to expressions too long to be reported here.

\subsection{Rule (v)}

Ref.~\cite{Gunter} studies a particle exclusion market model where the
order diffuse toward best prices and the rates are changed  according
to rule (v). According to the previous section, the 
trend duration pdf in this model is given by Eq~\req{F0t} (see also
Fig.~\ref{flight});  in
addition, the rates are tuned so as to ensure that $v=0$. Therefore
the trend duration has no definite average, which explains intuitively why without
order cancellation the price is always over-diffusive in this model; on the other hand,
when order cancellation is allowed, the effective rates have to be corrected;
this yields an effective $|v|=2\eta>0$, which is the origin of
the crossover to diffusive prices observed in this case.
Because this rule does not allow spread bouncing, the absolute value
of price increments is 0 or 1, and the sign of the increment stays
constant for a duration $F(l)$ that is given by
Eq~\req{F0t}. According to Ref.~\cite{Godreche}, the increment autocorrelation
$C(\tau)$ of such processes is the inverse Laplace transform of
\be
\hat C(s)=\frac{1}{s}\cro{1-2\frac{1-\hat F(s)}{s\avg{l}(1+\hat F(s))}}
\ee
where $F(l)$ is the pdf of trend duration $l$. Unfortunately, $\hat
C(s)$ cannot be inverted for $F$ given by Eq \req{F0t}. However, one can argue from this equation
 that for $\tau$ small enough, $C(\tau)\sim\tau^{-1/2}$, following Ref.~\cite{BouchMG}.
Another possibility is to consider a standard biased RW starting from
$s_0$ and
compute the probability that $s(t)<0$, neglecting the effects of the reflective boundary, which leads to
\be\label{autocorrG}
\avg{r(t)r(t+\tau)}=1-2P[r(t)=-r(t+\tau)]\simeq\erf\cro{\frac{v\tau+s_0}{\sqrt{2D\tau}}}
\ee
where the last approximation is valid as long as $v\tau+s_0\gg0$, that is,
as long as the spread is not likely to have reached the reflective boundary. 
When $\tau$ is such that $\frac{v\tau+s_0}{\sqrt{2Dt}}\ll 1$, that is
$\tau\ll 2D/|v|$, $\avg{r(t)r(t+\tau)}\propto \tau^{-1/2}$, with a
cutoff at $t\sim 2D/|v|$. Note that this result is exact if $v=0$, and gives $\avg{r(t)r(t+\tau)}\sim
\tau^{-1/2}$ for $\tau\gg s_0^2/2D$, in agreement with
Ref~\cite{Gunter}. The problem of this approach is that it gives
positive autocorrelation for all times, whereas it is negative at short times in real markets~\cite{Daco}.

\section{Price increment distribution}

All rules, except rule (v), lead to the same kind of price increment
pdf, which is made up of one part that is due to consecutive market orders on the
same side, and another that is due to spread bouncing (the price pdf
produced by rule (v) lacks the spread bouncing part). This implies
that when the rates are constant, the tails of $P(r)$ are given by that
of the spread pdf, which decreases exponentially in that case. In other
models, found e.g. in Refs~\cite{Maslov,CS01}, the pdf of the spread is
fat-tailed, hence that of the prices as well, which is arguably
artificial, since the mid-price increments $m(t)$ are also fat tailed
in real markets~\cite{Daco}.

It is possible to obtain a more refined pdf of
the trends. Up to here we essentially considered constant rates, which
lead to exponential tails. Time varying rates is the key to obtain fat tailed
return pdf in these models:
the mid-spread point $m$ moves indeed at average speed
$\frac{(\delta_B-\alpha_B)+(\alpha_A-\delta_A)}{2}=(\alpha_0+\alpha_1+\delta_0+\delta_1)\sigma$.
The rates are in
principle not bounded, as it is easy to generalise our model to
include multiple order depositions/annihilations during a time
step. In this case, they indicate the average number of events during
$\Delta t=1$ of a Poisson process. Their
distribution is related: about 10\% of deposited orders are matched in
the London Stock Exchange \cite{Coppejans} as well as in the ISLAND
ECN (www.island.com), a subpart of the NASDAQ~\cite{CS01}, hence one
can consider the deposition rate as
proportional to the annihilation rate in a first
approximation. $P(r)$ has fat tails as soon as $P(\alpha)\sim \alpha^{-\gamma}$. According to
Ref.~\cite{Stanleyvol}, this is the case, with an exponent of about
$3.5$. In addition, the autocorrelation of all the rates was found to
be algebraically decaying \cite{CS01}, which also shows up in that of the
number of transaction in a given time interval \cite{Stanleyvol}. We
shall explore further this generalisation in a forthcoming publication.

\section{Average densities}

\begin{figure}
\centerline{\includegraphics[width=8cm]{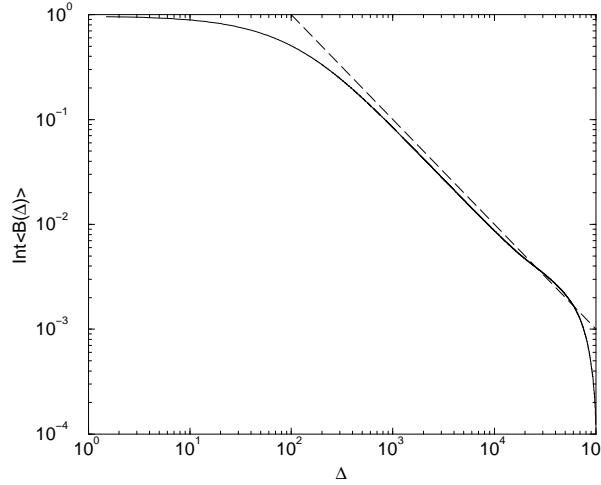}}
\caption{Integrated  average bid density
$\avg{B}(\Delta)$ versus $\Delta$, the position relative to the best bid for
$\eta=0.01$ and spread feed-back ($\delta=0.5$ $\alpha=0$, $c_0=2$, $c_1=3$). 
The dashed line has an exponent of -1.}
\label{figDens}
\end{figure}

Two recent papers \cite{FarmerLimit,BouchaudLimit} have focused 
on the average densities of orders relative to the best prices. In
both cases, these were found to have power law tails. The model of
Ref.~\cite{CS01} and its exclusion version, i.e. the model proposed here
with spread-feedback, are able to reproduce this behaviour (see fig
\ref{figDens});\footnote{One can even simplify further the exclusion
model by replacing one of the two order distributions by a wall, still
obtaining this result.} $\avg{B(\Delta)}$ is very well fitted with
the functional shape $(1+ax)^b$ with
$b\simeq-2.03$, very close to $-2$. This gives a market impact
exponent of $1/2$, which was measured in real markets, as reported Ref~\cite{FarmerImpact}. 
Note that the model of Ref.~\cite{FarmerLimit} also gives
the same exponent: this model can be seen as a special case $c_1=1/\sqrt{2}$,
$c_0=0$ of Ref.~\cite{CS01}, with a flat spatial  probability
distribution of
deposition; in particular, spread-feedback is present ($c_1>0$), which
seems to be necessary for obtaining densities with power-law tails.

Interestingly,
the functional shape of the average densities is the same as that of
the price in these models. In fact, it is also the same
as that of the spread, a general feature of market models with constant
rates, as explained above.  This last feature is probably very unrealistic, as
the pdf of the spread is most likely exponential in real markets. Nevertheless, this
can be easily explained. Suppose that the spread pdf has power tails,
then order deposition pdf has fat tails, hence, according to the
argument developed in ref~\cite{BouchaudLimit},
the average densities have power-law tails. In addition, as before, when the rates are constant, 
the tails of the price pdf are due to spread bouncing, hence are also
fat. In conclusion, in this type of models,
the spread, the price and the densities have all the same type of
pdf. But as argued above, fat tailed return should be obtained by
modulating the rates. Note that the model of Ref.~\cite{BouchaudLimit}
produces power-law tails for the densities by considering power-law
distributed relative positions for bulk deposition, which is
consistent with empirical findings.

\section{Conclusions}

The simplest particles models of limit order markets are exclusion
models and provide an ideal framework for mathematical analysis.
 We argued that particle models of limit order market should include
time-varying rates, and that the balance of rate only holds
on average, and is a crucial feature of these markets, explaining mid-term over-diffusive price
behaviour. We introduced several rules for time varying order deposition
and market order rates and  characterised the bid-ask
spread properties, computed Hurst plots and
correlation functions. The effects of order cancellation and
more general probability distributions of rates will be addressed in the future.

This work was supported by EPSRC under Oxford Condensed Matter Theory
grant GR/M04426. We thank G. M. Sch\"utz for seminal discussions.\\

\end{document}